\documentclass[referee]{raa}            

\usepackage{graphicx,times}             

\begin{document}

   \title{Does the gamma-ray signal from the central Milky Way indicate Sommerfeld enhancement of dark matter annihilation?}

   \volnopage{Vol.0 (200x) No.0, 000--000}      
   \setcounter{page}{1}          

   \author{Man Ho Chan}

   \institute{Department of Science and Environmental Studies, The Education University of Hong Kong \\
Tai Po, New Territories, Hong Kong, China}

   \date{Received~~2009 month day; accepted~~2009~~month day}

\abstract{Recently, Daylan et al. (2014) show that the GeV gamma-ray excess signal from the central Milky Way can be explained by the annihilation of $\sim 40$ GeV dark matter through $b\bar{b}$ channel. Based on the morphology of the gamma-ray flux, the best-fit inner slope of the dark matter density profile is $\gamma=1.26$. However, recent analyses of Milky Way dark matter profile favor $\gamma=0.6-0.8$. In this article, we show that the GeV gamma-ray excess can also be explained by the Sommerfeld-enhanced dark matter annihilation through $b\bar{b}$ channel with $\gamma=0.85-1.05$. We constrain the parameters of the Sommerfeld-enhanced annihilation by using the data from Fermi-LAT. We also show that the predicted gamma-ray fluxes emitted from dwarf galaxies generally satisfy the recent upper limits of gamma-ray fluxes detected by Fermi-LAT.
}

   \authorrunning{Man Ho Chan}            
   \titlerunning{Sommerfeld enhancement of dark matter annihilation}  

   \maketitle

%
%
\section{Introduction}           
\label{sect:intro}

In the past few years, some excess GeV gamma rays emitted from our Galactic center were reported (\cite{Hooper,Huang,Fermi}). The large diffuse signal of GeV gamma rays is hard to be explained by the cosmic ray and pulsar emission. Recent studies point out that the millisecond pulsars can only account no more than 10\% of the GeV excess (\cite{Hooper2,Daylan}). Therefore, the possibility of the emission of gamma rays due to the dark matter annihilation has become a hot topic in the recent years (\cite{Daylan,Gordon,Abazajian,Izaguirre,Calore}). 

In particular, \cite{Daylan,Calore} discover that the gamma-ray spectrum obtained from Fermi-LAT can be well fitted with $m=30-70$ GeV dark matter annihilation through $b\bar{b}$ channel. The cross section obtained $<\sigma v>=(1.4-2.0) \times 10^{-26}$ cm$^3$ s$^{-1}$ generally agrees with the expected canonical thermal relic abundance cross section $<\sigma v> \approx (2-3) \times 10^{-26}$ cm$^3$ s$^{-1}$. Moreover, the inner slope of the radial-dependence of the gamma-ray emission is $\gamma \approx 1.1-1.3$ (the best-fit value is $\gamma=1.26$), which is consistent with the theoretical expectation from numerical simulations ($\gamma=1-1.5$) (\cite{Daylan}). This work is further supported by a later study which includes the consideration of foreground and background uncertainties (\cite{Calore}). On the other hand, \cite{Chan} shows that this dark matter model can also explain the origin of hot gas near the Galactic center. Therefore, this dark matter model becomes one of the most popular models in dark matter astrophysics.

Besides the detection of gamma-ray emission from Galactic center, Fermi-LAT also obtains some upper limits of gamma-ray emission from dwarf galaxies and galaxy clusters. If we assume that the gamma-ray emission is due to the annihilation of $m=40$ GeV dark matter with $b\bar{b}$ channel, the corresponding upper limits of cross sections are $<\sigma v> \approx 1 \times 10^{-26}$ cm$^3$ s$^{-1}$ (\cite{Ackermann,Ackermann2}) and $<\sigma v> \approx (2-3) \times 10^{-25}$ cm$^3$ s$^{-1}$ (\cite{Ando}) for dwarf galaxies and galaxy clusters respectively. 

In fact, the results obtained in \cite{Daylan,Calore} assume that the annihilation cross section is constant (velocity-independent). However, it has been suggested that the annihilation cross section can be velocity-dependent. For example, the multiple exchange of some light force-carrier particle between the annihilating dark matter particle (the Sommerfeld enhancement) gives $<\sigma v> \propto v^{-\alpha}$, where $\alpha=1$ and $\alpha=2$ for non-resonance and resonance respectively (\cite{Sommerfeld,Zavala,Yang}). Furthermore, the inner slope of dark matter in our Galactic center revealed in \cite{Daylan,Calore} (best-fit $\gamma=1.26$) is a bit too large, compared with the recent observations in Milky Way (\cite{Pato}). Recent studies point out that the Milky Way dark matter density is well-fitted by a NFW density profile ($\gamma=1$) (\cite{Navarro,Iocco,Pato}). Detailed analyses in \cite{Pato} show that the best-fit $2\sigma$ range of the inner slope for the most representative baryonic model is $\gamma=0.6-0.8$. If we assume a generalized NFW profile with local density $\rho_{\odot}=0.4$ GeV cm$^{-3}$, $\gamma>1.2$ is excluded (outside the $5\sigma$ region) for this representative baryonic model. Although these results do not really rule out the possibility of having $\gamma=1.1-1.3$ (some baryonic models can still generate these values), such a large inner slope in Milky Way is certainly questionable. In fact, most of the inner slopes of dark matter density profiles observed do not show $\gamma>1$. For example, most galaxy clusters give $\gamma \approx 1$ (\cite{Pointecouteau}) and most galaxies and dwarf galaxies give $\gamma \le 1$ (\cite{Salucci2,Oh,Loeb}). Furthermore, recent numerical simulations show that baryonic feedback can decrease the inner slope of dark matter such that $\gamma<1$ for normal galaxies (\cite{Governato,Pontzen}). Therefore, the inner slope obtained in \cite{Daylan,Calore} does not give a good agreement with many other observations and recent numerical simulations.

In this article, we show that the result obtained in \cite{Daylan} is completely compatible with a velocity-dependent annihilation cross section. If we assume that the dark matter annihilation in the Milky Way center is Sommerfeld-enhanced, the resulting inner slope $\gamma$ obtained would give $\gamma=0.85-1.05$, which agrees with the standard NFW profile ($\gamma=1$). Also, we show that our model satisfies the Fermi-LAT results of nearby dwarf galaxies.


\section{The possibility of the Sommerfeld enhancement}
The observed gamma-ray flux within a solid angle $\Delta \Omega$ due to dark matter annihilation can be calculated by:
\begin{equation}
\phi(\Delta \Omega)=\frac{<\sigma v>}{8\pi m^2} \int \frac{dN_{\gamma}}{dE}dE \int_{\Delta \Omega}d \Omega \int_{\rm los} \rho^2 ds,
\end{equation}
where $dN_{\gamma}/dE$ is the photon spectrum per one dark matter annihilation (see Fig.~1) (\cite{Cembranos}) and $\rho$ is the dark matter density. The above equation is usually expressed as $\phi=\phi_{pp}J$, where 
\begin{equation}
\phi_{pp}=\frac{<\sigma v>}{8\pi m^2}\int \frac{dN_{\gamma}}{dE}dE
\end{equation}
is known as the `particle-physics factor' and
\begin{equation}
J=\int_{\Delta \Omega}d\Omega \int_{\rm los} \rho^2ds
\end{equation}
is known as the J-factor. 

Generally speaking, the best-fit annihilation channel, rest mass and the annihilation cross section of dark matter particle can be determined by the value of $\phi$ and the observed energy spectrum. In the above expressions, the annihilation cross section is assumed to be a constant. Therefore, only the integrand of the J-factor depends on $r$. However, if the annihilation cross section is velocity-dependent, Eq.~(1) has to be revised to include the effect of dark matter velocity.

Assume that the annihilation cross section is generally given by $<\sigma v>=<\sigma v>_0(v_0/v)^{\alpha}$, where $v$ is the velocity dispersion of dark matter particles, and $v_0$ and $<\sigma v>_0$ are constant. For the Sommerfeld enhancement, we have $\alpha \approx 1$ or $\alpha \approx 2$ for non-resonance and resonance cases respectively (\cite{Yang}). By putting the velocity-dependent cross section into Eq.~(1), we can rewrite the particle-physics factor and J-factor respectively as
\begin{equation}
\phi_{pp}'=\frac{<\sigma v>_0}{8\pi m^2}\int \frac{dN_{\gamma}}{dE}dE
\end{equation}
and
\begin{equation}
J'=\int_{\Delta \Omega}d\Omega \int_{\rm los} \rho^2 \left(\frac{v_0}{v} \right)^{\alpha}ds.
\end{equation}
Recent observations indicate that the dark matter density profile in our Galaxy is very close to an NFW profile (\cite{Iocco}). Therefore, for small $r$ region ($r \le 5$ kpc), we assume that $\rho= \rho_s(r/r_s)^{-1}$, where $\rho_s$ and $r_s$ are the scale density and scale radius respectively. The mass profile of the dark matter halo is $M_d=2 \pi \rho_sr_sr^2$. Since dark matter forms structure earlier than baryons, in an equilibrium configuration, the velocity dispersion of dark matter follows $v=C \sqrt{GM_d/r}$, where $C \sim 1$ is a constant which depends on the structure of the halo (\cite{Nesti}). However, most of the data of the observed morphology are obtained in the region $r \sim 1$ kpc (\cite{Calore}). When baryons collapse and form structures, the total mass of this small region would be dominated by the bulge mass. Although we know that the infall of baryons and some of the baryonic processes would affect the density distribution of dark matter, it is still not very clear quantitatively how it affects the dark matter distribution and the velocity distribution in our Galaxy. Recent numerical simulations point out that this effect is mainly determined by the ratio of stellar mass to the dark halo mass. Based on the study in \cite{Cintio}, the dark matter distribution of a Milky Way-size galaxy approaches the NFW profile for $r \sim 1$ kpc. However, based on the consideration of angular momentum, the velocity of dark matter particles would still change due to baryonic infall (adiabatic contraction). This would change the anisotropy coefficient $\beta$, which depends on the ratio of the tangential and radial velocities of dark matter particles. The analytic calculations in \cite{Vasiliev} show that if the change of the anisotropy coefficient is not very large, in some models, the velocity dispersion still follows the original NFW velocity distribution profile $v \propto \sqrt{r}$. In the following, although the mass profile is dominated by the bulge mass at $r \sim 1$ kpc, we assume that the distribution of dark matter and its velocity dispersion approximately follows the original NFW profile. In other words, we assume that the velocity distribution of dark matter is the same as the dark matter-only case, i.e. $v=C\sqrt{2\pi G\rho_sr_sr}$ \footnote{Note that here we did not fully consider the effect of baryons for the velocity distribution of dark matter. It is because the inclusion of baryons will not only affect the anisotropy but also the radial dependence of velocity distribution of dark matter.}. Even if the resulting velocity distribution deviates significantly from the original one, the numerical factor $C$ can also reflect some of the deviation in the following calculations. Based on the above assumption, we have
\begin{equation}
J'=\int_{\Delta \Omega}d\Omega \int_{\rm los} \rho^2 \left(\frac{r_0}{r} \right)^{\alpha/2}ds,
\end{equation}
where $r_0=v_0^2/2 \pi GC^2 \rho_sr_s$.

Since the Sommerfeld enhancement does not affect the energy spectrum of gamma rays $dN_{\gamma}/dE$, the best-fit annihilation channel and rest mass of dark matter particles in \cite{Daylan} would not be changed. The only parameter changed is the annihilation cross section. Based on the total gamma-ray flux detected, the velocity-dependent cross section can be constrained by
\begin{equation}
\phi=\phi_{pp}J=\phi_{pp}'J'.
\end{equation}

Technically, the Fermi-LAT observation is able to express the emission of gamma-ray flux as a function of $r$ ($F(r) \propto r^{-2 \gamma}$) (\cite{Daylan}). This function is directly proportional to the integrand of $J$. Therefore, to be consistent with the observed morphology, the integrand of $J$ and $J'$ must have the same $r$-dependence:
\begin{equation}
\rho_s^2 \left( \frac{r_s}{r} \right)^2 \left( \frac{r_0}{r} \right)^{\alpha/2}=\rho_s'^2 \left( \frac{r_s}{r} \right)^{2 \gamma},
\end{equation}
where $\rho_s'$ is the scale density used in \cite{Daylan}. By using the best-fit value $\gamma=1.26$ (\cite{Daylan}), we get $\alpha=2(2\gamma-2)=1.04 \approx 1$. In other words, the result obtained in \cite{Daylan} can also be interpreted as a non-resonant Sommerfeld-enhanced dark matter annihilation with an NFW dark matter density profile ($\gamma=1$). If we fix $\alpha=1$ and release the inner slope of dark matter density $\gamma$ to be a free parameter, the morphology of the gamma-ray flux $F(r) \propto r^{-(2.2-2.6)}$ from observations (\cite{Daylan}) would give $\gamma=0.85-1.05$, which is close to the $2 \sigma$ region ($\gamma=0.4-1$) for the most representative baryonic model (\cite{Pato}).

Based on the above formalism, we can obtain the value of $<\sigma v>_0v_0$ by using the flux $\phi$. By assuming $C=1$, $\gamma=1$, $r_s=20$ kpc and the local dark matter density $\rho_{\odot}=0.4$ GeV cm$^{-3}$ (\cite{Iocco}), we get $<\sigma v>_0v_0 \approx (2.2-3.2) \times 10^{-19}$ cm$^4$ s$^{-2}$. 

\begin{figure}
\vskip 5mm
 \includegraphics[width=82mm]{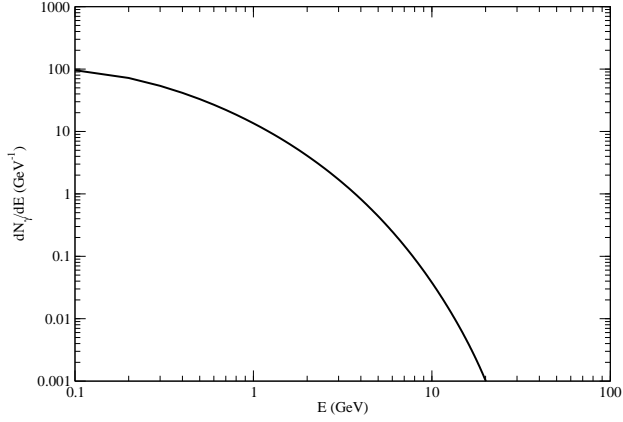}
 \caption{The photon spectrum per one dark matter annihilation through $b\bar{b}$ channel (\cite{Cembranos}). Here, we assume $m=40$ GeV.}
\vskip 5mm
\end{figure}

\section{Sommerfeld enhancement in dwarf galaxies}
If the Sommerfeld enhancement of dark matter annihilation occurs in our Galaxy, the same situation would occur in dwarf galaxies as well. In fact, the velocity dispersion near the center of a dwarf galaxy is small ($v \sim 10$ km/s). This small velocity dispersion would give a large annihilation rate near the centers of dwarf galaxies. 

From the result in \cite{Ackermann2}, the upper limit of annihilation cross section for the $b\bar{b}$ channel from the stacked analysis is $<\sigma v> \sim 2 \times 10^{-26}$ cm$^3$ s$^{-1}$ for $m \le 70$ GeV. However, if we include the effect of the Sommerfeld enhancement, we need to replace the J-factors used in \cite{Ackermann,Ackermann2} by Eq.~(6). Moreover, the J-factors used in \cite{Ackermann,Ackermann2} are somewhat larger than that in the recent empirical fits (\cite{Bonnivard,Evans}). Therefore, in the following, we use the lower limit of the J-factor for each dwarf galaxy in \cite{Bonnivard,Evans} and calculate a conservative upper limit of $<\sigma v>_0v_0$ based on the likelihood analysis in \cite{Ackermann2}.

Following \cite{Evans}, the J-factor of a dwarf galaxy can be given by an analytic formula:
\begin{equation}
J(\gamma)=\frac{25 \sigma_{los}^4}{64G^2} \frac{1}{D^2R_h} \left( \frac{D \theta}{R_h} \right)^{3-2\gamma}P(\gamma),
\end{equation}
where 
\begin{equation}
P(\gamma)=\frac{2}{\pi^{1/2}} \frac{(3-\gamma)^2 \Gamma(\gamma-0.5)}{(3-2\gamma) \Gamma(\gamma)},
\end{equation}
$\sigma_{los}$ is the velocity dispersion, $D$ is the distance of the galaxy, $\theta=0.5^{\circ}$ is the angular size and $R_h$ is the projected half-light radius (\cite{Evans}). By using Eq.~(6) and assuming $\gamma=1$ and $\alpha=1$, the revised J-factor for the Sommerfeld enhancement is given by
\begin{equation}
J'=J(1) \left(\frac{r_0}{D\theta} \right)^{1/2} \frac{P(1.25)}{P(1)}.
\end{equation}

By using the lower limits of the J-factor obtained in \cite{Bonnivard,Evans}, the combined revised J-factor is 2.3 times the original combined J-factor obtained in \cite{Ackermann2} (see Table 1 for the Sommerfeld-enhanced lower limit of the J-factor for each dwarf galaxy). Therefore, the results in \cite{Ackermann2} would give the upper limit of $<\sigma v>_0v_0$ to be $2 \times 10^{-19}$ cm$^4$ s$^{-2}$, which is close to our range $<\sigma v>_0v_0=(2.2-3.2) \times 10^{-19}$ cm$^4$ s$^{-2}$. Nevertheless, recent studies suggest that the dark matter density profiles of the dwarf galaxies are cored profiles ($\gamma<1$) instead of $\gamma=1$ (\cite{deBlok,Spekkens,Oh,Burkert2}). Therefore, the upper limit would be a bit larger because we have used the NFW profile ($\gamma=1$) to model the dark matter density profile of the dwarf galaxies. As estimated by \cite{Ackermann2}, this can give a factor of 1.3 in the upper limit. Also, if the inner density profiles of the dwarf galaxies are cored, the value of $C$ in dwarf galaxies may be a factor of 2 larger than that of Milky Way (\cite{Nesti}). As a result, if we include all the above mentioned factors, the corresponding upper limit of $<\sigma v>_0v_0$ would be larger by a factor of $1.8$ (i.e. $<\sigma v>_0v_0 \le 3.6 \times 10^{-19}$ cm$^4$ s$^{-2}$) and it would satisfy with the range observed by the Fermi-LAT for Milky Way. It means that the Sommerfeld-enhanced gamma-ray flux in dwarf galaxies does not exceed the observed upper limit.

\begin{table}
\caption{The lower limits of the Sommerfeld-enhanced J-factors $J'$ for different dwarf galaxies.}
 \label{table1}
 \begin{tabular}{@{}lc}
  \hline
  Dwarf galaxy & $\log(J'/\rm GeV^2~cm^{-5})$  \\
  \hline
  Bootes I & 17.0 \\
  Canes Venatici II & 17.6 \\
  Carina & 18.2 \\
  Coma Berenices & 19.1 \\
  Draco & 19.3 \\
  Fornax & 18.2 \\
  Hercules & 16.9 \\
  Leo II & 17.6 \\
  Leo IV & 13.4 \\
  Sculptor & 19.0 \\
  Segue I & 13.5 \\
  Sextans & 17.9 \\
  Ursa Major II & 19.8 \\
  Ursa Minor & 19.2 \\
  Willman 1 & 18.5 \\
  Canes Venatici I & 17.6 \\
  Leo I & 17.8 \\
  Ursa Major I & 18.7 \\
  \hline
 \end{tabular}
\end{table}

\section{Discussion}
Previously, \cite{Daylan} show that the GeV gamma-ray excess can be explained by the annihilation of $\sim 40$ GeV dark matter through $b\bar{b}$ channel. Based on the morphology of the gamma-ray flux, the best-fit inner slope of dark matter density profile is $\gamma=1.26$. However, recent analyses show that the best-fit $2\sigma$ range of inner slope for most the representative baryonic model is $\gamma=0.6-0.8$ (\cite{Pato}). Also, many observations indicate $\sigma \le 1$ (\cite{Salucci2,Oh,Loeb}). In this article, we show that the GeV gamma-ray excess can also be explained by the Sommerfeld-enhanced dark matter annihilation through $b\bar{b}$ channel with $\gamma=1$ (the NFW profile). In general, our model is compatible with the range of the inner slope $\gamma=0.85-1.05$. By using the results in \cite{Daylan}, we also constrain the parameters of the Sommerfeld enhancement: $\alpha=1$ and $<\sigma v>_0v_0 \approx (2.2-3.2) \times 10^{-19}$ cm$^4$ s$^{-1}$ for $\gamma=1$. 

Although the annihilation model with Sommerfeld enhanced cross section is more complicated, this model can fully explain the morphology of the gamma-ray flux and favor the smaller inner slope of the dark matter density in Milky Way. Since the morphology of the gamma-ray flux gives $F \propto r^{-(2.2-2.6)}$ (\cite{Daylan}), the inner slope obtained from this model is $\gamma=0.85-1.05$, which gives a very good agreement with the recent analysis in \cite{Pato}. However, if we assume the constant cross section for dark matter annihilation, the required inner slope is $\gamma=1.1-1.3$, which does not satisfy with the observed $2\sigma$ range of the inner slope $\gamma=0.6-0.8$ for the most representative baryonic model (\cite{Pato}). Therefore, our model can alleviate the tension between the existing dark matter annihilation model and the observations.

In fact, the Sommerfeld enhancement would greatly enhance the dark matter annihilation rate near the dwarf galactic center because the velocity dispersion is very small there. Therefore, we predict that a strong signal of gamma-ray flux at the dwarf galactic center would be resulted if our model is correct. We show that the gamma-ray fluxes emitted due to the Sommerfeld-enhanced dark matter annihilation from the dwarf galaxies generally satisfy the current upper limits obtained by the 6-year Fermi-LAT data. If the Fermi-LAT can further constrain the upper limits of the gamma-ray flux or the detected gamma-ray spectrum in the future, we can get a tighter constraint on the annihilation cross section as well as the dark matter rest mass.

\begin{acknowledgements}
I am grateful to the referee for helpful comments on the manuscript. This work is partially supported by a grant from The Education University of Hong Kong (Project No.:RG57/2015-2016R).
\end{acknowledgements}

\label{lastpage}

\end{document}